\documentclass[aps,preprint,superscriptaddress]{revtex4-1}

\usepackage{amsmath}
\usepackage{fixmath}
\usepackage{accents}
\usepackage{graphicx}%

\usepackage{hyperref}

\newcommand{\mb}{\mathbold}

\begin{document}
\title{Purcell enhancement of the parametric down-conversion in two-dimensional nonlinear materials}
\author{Mikhail Tokman}
\affiliation{Institute of Applied Physics, Russian Academy of Sciences, 603950 Nizhny Novgorod, Russia}
\author{Zhongqu Long}
\affiliation{Department of Physics and Astronomy, Texas A\&M
University, College Station, TX, 77843 USA}
\author{Sultan AlMutairi}
\affiliation{Department of Physics and Astronomy, Texas A\&M
University, College Station, TX, 77843 USA}
\author{Yongrui Wang}
\affiliation{Department of Physics and Astronomy, Texas A\&M
University, College Station, TX, 77843 USA}
\author{Valery Vdovin}
\affiliation{Institute of Applied Physics, Russian Academy of Sciences, 603950 Nizhny Novgorod, Russia}
\author{Mikhail Belkin}
\affiliation{Department of Electrical and Computer Engineering, 
University of Texas at Austin, Austin, TX, 78712 USA}
\author{Alexey Belyanin}
\affiliation{Department of Physics and Astronomy, Texas A\&M
University, College Station, TX, 77843 USA}

\date{\today}

\begin{abstract} 

Ultracompact nonlinear optical devices utilizing two-dimensional (2D) materials and nanostructures are emerging as important elements of photonic circuits. Integration of the nonlinear material into a subwavelength cavity or waveguide leads to a strong Purcell enhancement of the nonlinear processes and compensates for a small interaction volume. The generic feature of such devices which makes them especially challenging for analysis is strong dissipation of both the nonlinear polarization and highly confined modes of a subwavelength cavity. Here we solve a  quantum-electrodynamic problem of the spontaneous and stimulated parametric down-conversion in a nonlinear quasi-2D waveguide or cavity. We develop a rigorous Heisenberg-Langevin approach which includes dissipation and fluctuations in the electron ensemble and in the electromagnetic field of a cavity on equal footing. Within a relatively simple model, we take into account the nonlinear coupling of the quantized cavity modes, their interaction with a dissipative reservoir and the outside world, amplification of thermal noise and zero-point fluctuations of the electromagnetic field, and other relevant effects. We derive closed-form analytic results for relevant quantities such as the spontaneous parametric signal power and the threshold for parametric instability. We find a strong reduction in the parametric instability threshold for 2D nonlinear materials in a subwavelength cavity and provide a comparison with conventional nonlinear photonic devices. 

\end{abstract}

\maketitle
\section{Introduction}
Enhancement of the radiative processes due to the localization of emitters in a subwavelength cavity (so-called Purcell enhancement \cite{purcell}) is a fundamental cavity quantum electrodynamics (QED) effect of great importance for numerous applications. The bulk of the research has been focused on exploring the enhancement of spontaneous emission in various compact radiation sources from single quantum emitters to LEDs and nanolasers. The nonlinear optics has received relatively less attention; however, recent advancements in strong light localization using subwavelength cavities, photonic crystals, metamaterials, and metasurfaces enabled the nonlinear optics in ultrasmall volumes and at relatively low power levels; see e.g. \cite{lin1994,bermel2007,genevet2010,kauranen2012,kivshar2014,lee2014,lee2016,smirnova2016,atwater2017} and references therein. The rise of 2D materials with atomic monolayer thickness  and excellent nonlinear optical properties, such as graphene \cite{hendry2010,yao2012,yao2014,wang2016} and transition metal dichalcogenide monolayers \cite{malard2013,liu2016} has enabled quasi-2D cavities and waveguides only a few nm thick \cite{koppens2016,ye2015}.  These advances create new exciting opportunities for ultracompact nonlinear optical devices, but also raise important  issues of the correct description of quantum fields in systems with  strong dissipation both in a macroscopic ensemble of fermionic emitters (e.g. a 2D semiconductor or monolayer graphene, or a 2D electron gas in a quantum well) and for the electromagnetic (EM) field in a cavity. 

One important application for Purcell-enhanced nonlinear optics is compact systems for generation of squeezed and entangled photon states as a result of parametric down-conversion. Such systems are inevitably lossy. The general approach to introducing dissipation and corresponding fluctuations has been known for a long time and is based on the Heisenberg-Langevin formalism; e.g. \cite{SZ,gardiner,et2017,david,belyanin2013}.  Its generalization to systems far from equilibrium, with arbitrary dissipation mechanisms and arbitrary photon density of states is nontrivial; see e.g. \cite{et2015,te2013}. Recent work \cite{atwater2017}  considered the process  of spontaneous parametric down-conversion in hyperbolic metamaterials, in which the EM field dissipation and fluctuations are due to an equilibrium thermal reservoir. In the present paper we consider both spontaneous and stimulated parametric down-conversion  in a generic quasi-2D subwavelength cavity, taking into account dissipation and fluctuations  both due to absorption in the intracavity material and due to in/outcoupling of the intracavity EM field with the outside world. 

We generalize the properties of Langevin noise sources known for a single mode of a quantized field (e.g. \cite{SZ,gardiner,tokman2018}) to an ensemble of coupled field oscillators. We derive the properties of the Langevin sources needed to conserve their commutation relations and show that they are not affected by a more complicated dynamics of coupled Heisenberg field operators; moreover, this statement does not depend on any specific microscopic model of a dissipative reservoir. We are able to derive closed-form analytic results for the spontaneous parametric signal, the parametric gain, and the threshold for parametric amplification. These expressions include the contributions from all relevant  dissipation and fluctuation effects such as absorption and radiation losses, interaction with thermal and zero-point fluctuations, parametric amplification of thermal noise and seed photons at the signal frequency, etc.; see e.g. Eqs.~(\ref{Eq:76}), (\ref{spont}) below. 

Our approach has obvious limitations of a Heisenberg-Langevin formalism, namely it assumes that  the coupling of a dynamic subsystem to a dissipative reservoir is sufficiently weak. If this is not the case and the coupling to other EM modes, photons, etc. is strong, one would have to include it as part of an ``exact'' Hamiltonian dynamics, in which case there would be no need in adding the corresponding Langevin sources and the commutation relations would be satisfied automatically.  We also do not investigate the nonlinear stage of parametric oscillations accompanied by the pump depletion,  nonlinear evolution of phasematching conditions,  nonlinear modification of refraction and diffraction losses, and other nonlinear effects that are essentially classical and depend on a particular experimental setup. 

Section II describes the spatial structure of the EM field in a subwavelength quasi-2D electrodynamic structure, develops   the quantization procedure in a dissipationless system, and discusses three-wave mode matching conditions. Section III introduces the Heisenberg-Langevin approach for the parametric down-conversion in a dissipative cavity. It derives convenient analytic expressions for the spontaneous parametric signal, the parametric amplification threshold in plane-parallel cavities,  and the signal evolution at the linear stage. We discuss several numerical examples for the parametric down-conversion in quasi-2D systems studied by other groups. Section IV compares parametric amplification threshold in a subwavelength cavity with the one in a standard Fabry-Perot cavity containing a 2D nonlinear layer.  In this case the performance tradeoff is between the cavity losses and the modal overlap with a nonlinear layer. Larger cavities tend to have a higher Q-factor but lower coupling to a nonlinear 2D layer.  Our results show that it is possible to achieve a significant  reduction of the parametric amplification threshold due to Purcell enhancement in quasi-2D subwavelength cavities.

\section{Parametric down-conversion in a conservative system }

Consider three cavity modes with frequencies related by the energy conservation in the parametric down-conversion process: 
\begin{equation}
\label{Eq:51}
	\omega_p=\omega_s+\omega_i.
\end{equation}
Here the pumping at frequency $\omega_p$ will be considered a classical coherent field, 
\begin{equation}
\label{Eq:52}
	\mb{E}_p=\mb{E}_p(\mb{r})e^{-i\omega_pt}+C.C.
\end{equation}
The field at signal and idler frequencies, $\omega_s$ and $\omega_i$ , will be the quantum field described by the operator 
\begin{equation}
\label{Eq:53}
	\mb{\hat{E}}=\sum_{\nu=s,i}\lbrack\mb{E}_\nu(\mb{r})\hat{c}_\nu+\mb{E}^*_\nu(\mb{r})\hat{c}^\dagger_\nu\rbrack,
\end{equation}
where $\hat{c}_\nu$ and $\hat{c_\nu}^\dagger$ are boson annihilation and creation operators. The functions $\mb{E}_{p,s,i}(\mb{r})$ in Eqs.~(\ref{Eq:52}) and (\ref{Eq:53}) are determined by the spatial structure of the cavity modes. 
The normalization of functions $\mb{E}_{\nu}(\mb{r})$ needs to be chosen in such a way that the commutation relation for boson operators $\hat{c}_\nu$ and $ \hat{c_\nu}^\dagger$ have a standard form $[\hat{c}_\nu,\hat{c_\nu}^\dagger]=\delta_{\nu\nu'}$. Following \cite{fain,vdovin,tokman2016}, one can obtain 
\begin{equation} 
\label{3a} 
\int_V E_{\nu j}^*(\mb{r}) \frac{1}{2 \omega_{\nu}} \left[ \frac{ \partial \left( \omega^2 \varepsilon_{jk}(\omega,\mb{r}) \right) }{\partial \omega} \right]_{\omega = \omega_{\nu}}  E_{\nu k}(\mb{r})\, d^3r = 2 \pi \hbar \omega_{\nu}, 
\end{equation} 
where $\omega_{\nu}$ is the eigenfrequency of a cavity mode, $E_{\nu j}(\mb{r}) $ is a Cartesian component of the vector field $\mb{E}_{\nu}(\mb{r})$, $\varepsilon_{jk}(\omega,\mb{r}) $ is the dielectric tensor, and $V$ is a cavity volume (a quantization volume). 

Equation (\ref{3a}) is valid when the dissipation is weak enough. Specifically, the following three conditions have to be satisfied for a dissipation rate $\Gamma$ of a given cavity mode.  The first condition is obvious: $\Gamma \ll \omega$ has to be true for the frequencies of all modes involved in the parametric process. The second condition implies that the change of the Hermitian dielectric function $\varepsilon_{jk}(\omega)$ has to be small over the frequency interval of the order of $\Gamma$:  $|(\partial \varepsilon_{jk}(\omega)/\partial \omega) \Gamma | \ll   |\varepsilon_{jk}(\omega)|$.  The third condition states that the change in the derivative of $ \varepsilon_{jk}(\omega)$ which enters the expression for the EM energy density in Eq.~(\ref{3a}) must also be small:  $|(\partial^2 \varepsilon_{jk}(\omega)/\partial\omega^2) \Gamma |\ll  |(\partial \varepsilon_{jk}(\omega)/\partial \omega)|$.

Consider a 3D cavity filled with an isotropic dielectric medium, as sketched in Fig.~1. The cavity thickness in $z$-direction is much smaller than wavelength: $ L_z \ll c/ \sqrt{\bar{\varepsilon}}\omega$ , where $\bar{\varepsilon}$ is a typical (average) value of the dielectric constant
 of the filling. As was shown in \cite{tokman2018}, if the dielectric filling consists of plane-parallel layers, i.e. $\varepsilon=\varepsilon(z)$, the structure of the cavity eigenmodes is quasi-electrostatic along the $z$-axis, i.e. $E_z \varepsilon(z) \approx$ const, $E_{x,y} \ll E_z$. In this case the field of a cavity mode can be written as 
 \begin{equation}
 \label{4} 
 \mb{E}_{p,s,i}(\mb{r}) \approx \mb{z}_0 D_{p,s,i} \frac{\zeta_{p,s,i} (x,y)}{\varepsilon(\omega_{p,s,i},z)},
 \end{equation}
 where the constants $D_{p,s,i}$ are coordinate-independent amplitudes of the electric induction. To find the functions $\zeta_{p,s,i} (x,y)$ we solve the wave equation 
 \begin{equation} \nonumber  
 \nabla \cdot \left( \nabla \cdot   \mb{E}_{p,s,i} \right) - \nabla^2  \mb{E}_{p,s,i} - \frac{\omega_{p,s,i}^2}{c^2} \varepsilon(\omega_{p,s,i},z) \mb{E}_{p,s,i} = 0. 
 \end{equation} 
 Consider a $z$-component of this equation,
 \begin{equation}
 \label{4a} 
 \frac{\partial}{\partial z} \left( \frac{\partial E_{(p,s,i)x}}{\partial x} +  \frac{\partial E_{(p,s,i)y}}{\partial y} \right) = \left[ \frac{\partial^2}{\partial x^2} + \frac{\partial^2}{\partial y^2} + \frac{\omega_{p,s,i}^2}{c^2} \varepsilon(\omega_{p,s,i},z) \right] E_{(p,s,i)z}. 
 \end{equation}
 
 Following the procedure described in \cite{tokman2018}, we integrate Eq.~(\ref{4a}) over $\int_{-L_z/2}^{L_z/2} dz$   
 taking into account the boundary conditions on the metal planes of the cavity, $E_{x,y}(\pm L_z/2) = 0$.  Then we substitute Eq.~(\ref{4}) into the result of integration, which gives
 \begin{equation}
 \label{5} 
 \left[ \frac{\partial^2}{\partial x^2} +  \frac{\partial^2}{\partial y^2} + \displaystyle \frac{\omega_{p,s,i}^2}{c^2 \frac{1}{L_z} \int_{-L_z/2}^{L_z/2} \frac{1}{\varepsilon(\omega_{p,s,i},z)} dz}  \right] \zeta_{p,s,i} = 0. 
 \end{equation}
 The solution to Eq.~(\ref{5}) with zero boundary conditions at the edges of the cavity  determines eigenfrequencies and the structure of the eigenmodes for a quasi-2D cavity with an arbitrary shape in the $(x,y)$-plane. 
 
 Similar equations can be derived if one simply utilizes jumps of the dielectric constants on the sides instead of metal coating. Even without any jump in the dielectric constants, an open end of a thin waveguide with vertical size much smaller than wavelength is  a good reflector and therefore any radiation losses through the facets are small and are not affecting the mode spatial structure significantly.
 
 The expression for the constants $D_{p,s,i}$ for quantized fields can be obtained from the general equation Eq.~(\ref{3a}) (see also \cite{tokman2018}),   
 \begin{equation}
 \label{6} 
|D_{\nu}|^2 =  \displaystyle \frac{2 \pi \hbar \omega_{\nu}}{\int_S |\zeta_{\nu}(x,y)|^2 d^2r   \frac{1}{2\omega_{\nu}} \int_{-L_z/2}^{L_z/2} \left[ \frac{\partial (\omega^2 \varepsilon(\omega,z))}{\partial \omega} \right]_{\omega = \omega_{\nu}} dz}, 
 \end{equation}
where $\nu = p,s,i$. For a simple case of a rectangular-shaped cavity, when $S = L_x \times L_y$, where $L_x$ and $L_y$ are the cavity dimensions along $x$ and $y$ directions, it is easy to obtain useful analytic expressions for the modal spatial structure and frequencies. For eigenmodes with one half-wavelength along the $y$-axis and  $N$ half-wavelengths along the $x$-axis, we obtain the modal profile 
 \begin{equation}
\label{Eq:new11}
\zeta_{\nu}=\cos \left(\frac{\pi y}{L_y}\right) \times \left\{  
\begin{matrix}
\cos\left(\displaystyle\frac{N_{odd}^{(\nu)}\pi x}{L_x}\right) \\ \sin \left(\displaystyle\frac{N_{even}^{(\nu)}\pi x}{L_x}\right)
\end{matrix} 
\right. ,
\end{equation}
\begin{equation}
\label{6c}
\int_S |\zeta_{\nu}(x,y)|^2 d^2r = \frac{S}{4}.
\end{equation}
The eigenfrequencies are given by 
\begin{equation}
\label{Eq:new10}
\left(\frac{N_{\nu}\pi}{L_x}\right)^2+\left(\frac{\pi}{L_y}\right)^2 = \frac{\omega_{\nu}^2}{c^2}\frac{L_z}{ \int\limits_{- \frac{L_z}{2}}^{+\frac{L_z}{2}} \frac{1}{\varepsilon(\omega_{\nu},z)} dz}.
\end{equation}
For a particular case of a uniform dielectric constant, Eqs.~(\ref{Eq:new11}) and (\ref{Eq:new10}) are exact, i.e. they do not require an assumption of a quasielectrostatic field structure.

\begin{figure}[h]
    \centering
    \includegraphics[width=1\columnwidth]{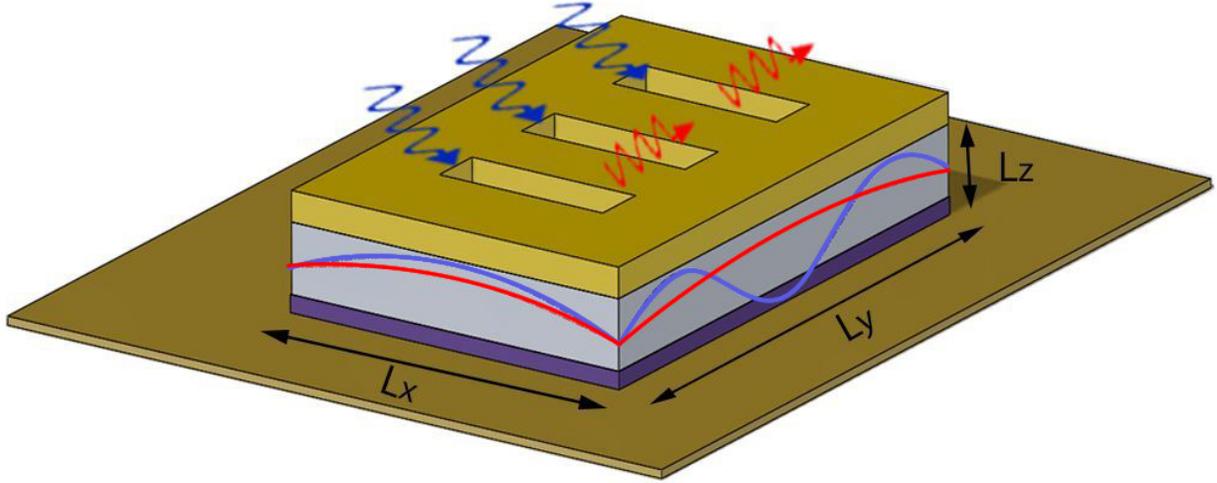}
    \caption{A sketch of a nanocavity with thickness $L_z$ much smaller than wavelength. The profiles of the electric fields  of the TE$_{013}$ pump mode (blue) and TE$_{011}$ signal and idler modes (red) are sketched on the sides. Dark blue layer indicates a 2D nonlinear material; light blue layer is a cavity filling. Top and bottom gold layers are metal plates. The radiation can be in/outcoupled through the gratings or cavity edges.  }
	\label{fig1}
\end{figure}

We assume that a 2D electron system with the second-order nonlinear susceptibility is positioned in the cavity. The material can be a quantum well (QW), a 2D semiconductor, and even graphene, which has a strong second-order nonlinearity beyond electric-dipole approximation, despite being centrosymmetric \cite{wang2016}.  The second-order nonlinearity gives rise to the nonlinear polarization at signal and idler frequencies. The excitation equations for the cavity modes derived from the operator-valued Maxwell's equations \cite{fain} take the form 
\begin{equation}
\label{Eq:54}
    \dot{\hat{c}}_\nu+i\omega_\nu\hat{c}_\nu=
-\frac{i}{\hbar\omega^2_\nu}\int\limits_V\ddot{\hat{\mb{P}}}_{NL}(\mb{r},t)\mb{E}^*_\nu(\mb{r})d^3r.
\end{equation}
The nonlinear polarization $\hat{\mb{P}}_{NL}(\mb{r},t)$ should be determined for a given electron system; in general, it has a nonuniform distribution over the cavity cross-section. However, it is obvious from Eq.~(\ref{Eq:54}) that only the integral over the nonlinear polarization matters. Therefore it is convenient to consider a nonlinear layer with uniform dielectric constant $\varepsilon_{QW}(\omega_{p,s,i})=\varepsilon_{p,s,i}$ and uniform second-order nonlinear susceptibilities $\chi^{(2)}(\omega_s=\omega_p-\omega_i)=\chi_s^{(2)}$, $\chi^{(2)}(\omega_i=\omega_p-\omega_s)=\chi_i^{(2)}$, $\chi^{(2)}(\omega_p=\omega_s+\omega_i)=\chi_p^{(2)}$. For a general case of a nonuniform layer the above quantities can be considered as parameters obtained as a result of integration in Eq.~(\ref{Eq:54}).

For a uniform layer the nonlinear polarization can be expressed as 
\begin{equation}
\label{Eq:57}
     \hat{\mb{P}}_{NL}=\mb{z}_0\zeta_p(x,y)\lbrack
\zeta_i(x,y)\chi_s^{(2)}E_p E^*_ie^{-i\omega_pt}\hat{c}^\dagger_i
+\zeta_s(x,y)\chi_i^{(2)}E_p E^*_se^{-i\omega_pt}c^\dagger_s
\rbrack+H.C.;
\end{equation}
where $c_s^\dagger\propto e^{i\omega_st}$, $c_i^\dagger\propto e^{i\omega_it}$, 
\begin{equation}
\label{Eq:56}
    |E_{s,i}|^2=\frac{8\pi\hbar\omega_{s,i}}{L_x L_y L_z\displaystyle\frac{1}{2\omega_{s,i}}
	\left[ \frac{\partial(\omega^2\epsilon)}{\partial\omega}\right]_{\omega=\omega_{s,i}}},
\end{equation}
and $E_p =$ const. In Eq.~(\ref{Eq:56}) we assumed a rectangular cavity shape for simplicity.

 If the nonlinearity is non-dissipative, the nonlinear susceptibilities satisfy the symmetry properties which ensure than Manley-Rowe relationships are satisfied in a conservative system \cite{keldysh, te2011}.
\begin{equation}
\label{Eq:58}
     \chi_s^{(2)}= \chi_i^{(2)}= \chi_p^{(2)*}= \chi^{(2)}.
\end{equation}
Using the rotating wave approximation, Eqs.~(\ref{Eq:54}) and (\ref{Eq:57}) give the following parametrically coupled equations for a given classical  pumping:  
\begin{equation}
\label{Eq:59}
     \dot{\hat{c}}_s+i\omega_s\hat{c}_s+\varsigma e^{-i\omega_pt}\hat{c}_i^\dagger=0,\hspace{5 mm}
          \dot{\hat{c}}_i^\dagger-i\omega_i\hat{c}^\dagger_i+\varsigma^* e^{+i\omega_pt}\hat{c}_s=0
\end{equation}
where
\begin{gather}
\label{Eq:60}
  \varsigma=-\frac{i}{\hbar}\chi^{(2)}lE_pE_i^*E_s^*\times J,
\\
\label{Eq:61}
J=\int\limits_{L_x\times L_y}\zeta_p(x,y)\zeta_i(x,y)\zeta_s(x,y)d^2r
\end{gather}
is a modal overlap factor and $l$ is the thickness of the nonlinear layer in $z$ direction.

Equations (\ref{Eq:59}) are Heisenberg equations for the effective Hamiltonian 
\begin{equation}
\label{Eq:62}
     \hat{H}=\hbar\omega_s\left(\hat{c}_s^\dagger \hat{c}_s+\frac{1}{2}\right)
+\hbar\omega_i\left(\hat{c}_i^\dagger \hat{c}_i+\frac{1}{2}\right)
-i\hbar\varsigma e^{-i\omega_pt}\hat{c}_s^\dagger\hat{c}_i^\dagger
+i\hbar\varsigma^* e^{i\omega_pt}\hat{c}_s\hat{c}_i.
\end{equation} 
For a parametric down-conversion of a pump photon into two identical photons, when 
\begin{equation}
\label{Eq:63}
     \omega_p=2\omega_s
\end{equation}
we arrive at the Hamiltonian
\begin{equation}
\label{Eq:64}
      \hat{H}=\hbar\omega_s\left(\hat{c}_s^\dagger \hat{c}_s+\frac{1}{2}\right)
-i\hbar\frac{\varsigma}{2} e^{-i\omega_pt}\hat{c}_s^\dagger\hat{c}_s^\dagger
+i\hbar\frac{\varsigma}{2}^* e^{i\omega_pt}\hat{c}_s\hat{c}_s.
\end{equation}

The condition $J \neq 0$ is similar to the three-wave phase matching condition for the wave vectors of modes participating in the parametric down-conversion. The phase matching needs to be satisfied together with frequency matching in Eq.~(\ref{Eq:51}), which could be highly nontrivial in a 3D geometry and in a dispersive medium. An important advantage of a subwavelength cavity is that these conditions are straightforward to satisfy by adjusting the cavity geometry. Indeed, consider the decay of the pump into two lowest-order TE$_{011}$ modes satisfying Eq.~(\ref{Eq:63}); in this case $\zeta_s(x,y)= \cos\left( \displaystyle\frac{\pi y}{L_y}\right) \cos\left( \displaystyle\frac{\pi x}{L_x}\right)$. For a pumping mode of TE$_{01N}$ type with N even, $J=0$; however for N odd we get $J\neq0$. For example, for a TE$_{013}$ pumping mode (see Fig.~1), when $     \zeta_p(x,y)= \cos\left( \displaystyle\frac{\pi y}{L_y}\right) \cos\left( \displaystyle\frac{3\pi x}{L_x}\right)$, we obtain: $     J=\displaystyle\frac{L_x L_y}{\pi^2}\times\frac{16}{45}$. In this case from Eq.~(\ref{Eq:63}) and mode frequencies given by Eq.~(\ref{Eq:new10}) one can get a condition for cavity sizes: 
\begin{equation}
\nonumber
     \frac{L_x}{L_y}=\sqrt{\frac{9-4\displaystyle\frac{\varepsilon(2\omega_s)}{\varepsilon(\omega_s)}}
	{4\displaystyle\frac{\varepsilon(2\omega_s)}{\varepsilon(\omega_s)}-1}}
\end{equation}

\section{Equations for parametric down-conversion in a dissipative system. Heisenberg-Langevin approach }

Here we take into account absorption and radiative losses within the Heisenberg-Langevin formalism. We remind the reader that this approach assumes that the coupling of a dynamic subsystem to a dissipative reservoir is sufficiently weak. If this is not the case and the coupling is strong, 
the process of energy loss by a given EM mode should be described within a closed Hamiltonian system (e.g. as a coupling to other EM modes, phonons etc.). In this case one does not need any Langevin sources, because in a Hamiltonian system proper commutation relations are satisfied automatically. Whenever the energy exchange of a dynamical subsystem with a reservoir is relatively weak and can be considered within a phenomenological approach, the  ``dissipation + the Langevin noise'' model should be valid for any mechanism of dissipation.  

For example, we assume here that the spatial mode structure corresponds to the one in an ideal cavity, whereas diffraction losses of the field out of a cavity can be described by an effective loss rate. This assumption obviously works as long as losses do not affect the mode structure significantly. If the latter is not true, one would have to solve a rigorous diffraction problem which couples the field modes in the cavity and outside. For such a rigorous problem all commutation relations would be satisfied automatically. 

 Introducing  operators of slowly varying field amplitudes, namely $\hat{c}_{s,i}=\hat{c}_{0s,i}(t)e^{-i\omega_{s,i}t}$, $\hat{c}_{s,i}^{\dagger}=\hat{c}^\dagger_{0s,i}(t)e^{+i\omega_{s,i}t}$, we obtain from Eqs.~(\ref{Eq:59}) the following equations:
	\begin{equation}
	\label{Eq:66}
\left.
\begin{array}{r}
\dot{\hat{c}}_{0s}+\Gamma_s\hat{c}_{0s}+\varsigma\hat{c}_{0i}^\dagger=\hat{L}_s \\
\dot{\hat{c}}_{0i}+\Gamma_i\hat{c}_{0i}+\varsigma\hat{c}_{0s}^\dagger=\hat{L}_i
\end{array} \right\},
\end{equation}
where $\Gamma_{s,i}=\Gamma_{r(s,i)}+\Gamma_{\sigma(s,i)}$, the coefficients $\Gamma_{r(s,i)}$ and $\Gamma_{\sigma(s,i)}$ denote, respectively, radiative losses due to the outcoupling of radiation from the cavity and absorption losses due to intracavity absorption.  $\hat{L}_{s,i}$ are the Langevin noise operators. We show in Appendix A  that to preserve commutation relations $[\hat{c}_{0i},\hat{c}_{0i}^\dagger]=[\hat{c}_{0s},\hat{c}_{0s}^\dagger]=1$ at $\Gamma_{s,i}\neq0$ the noise operators in the right-hand side of Eqs.~(\ref{Eq:66}) should satisfy the same commutation relations as in the case of one quantum oscillator \cite{SZ,gardiner,tokman2018} and they should also commute with each other: 
\begin{equation}
\label{Eq:67} 
	\left[ \hat{L}_s(t'),\hat{L}_s^\dagger(t)\right]=2\Gamma_s\delta(t-t'),\hspace{5 mm}
\left[ \hat{L}_i(t'),\hat{L}_i^\dagger(t)\right]=2\Gamma_i\delta(t-t'),\hspace{5 mm}
	\left[ \hat{L}_s(t'),\hat{L}_i^\dagger(t)\right]=0.
\end{equation}
The fact that commutation relations are the same for one quantum oscillator and for two (or more) interacting oscillators is expected, since the processes within the Hamiltonian dynamics do not affect the commutators; this can be easily checked, for example for the system described by the Hamiltonian Eq.~(\ref{Eq:62}). Noise correlators can be defined by generalizing the simplest expression in \cite{SZ} to the case of two absorbers with different temperatures: 
	\begin{equation}
	\label{Eq:68}
\left.
\begin{array}{r}
\langle\hat{L}^\dagger_s(t)\hat{L}_s(t')\rangle=2\left[\Gamma_{\sigma s}n_{T_\sigma}(\omega_s)+\Gamma_{r s}n_{T_r}(\omega_s)\right]\delta(t-t') \\
\langle\hat{L}^\dagger_i(t)\hat{L}_i(t')\rangle=2\left[\Gamma_{\sigma i}n_{T_\sigma}(\omega_i)+\Gamma_{r i}n_{T_r}(\omega_i)\right]\delta(t-t')
\end{array} \right\}
\end{equation}
where $n_{T_{r,\sigma}}$ is the average number of thermal photons  at temperature $T_{r,\sigma}$;  $T_r$ and $T_{\sigma}$ denote the temperature outside and inside the cavity, respectively. Expressions (\ref{Eq:68}) imply that dissipative and radiative noises are not correlated.

The solution to Eqs.~(\ref{Eq:66}) can be represented as \cite{et2017,belyanin2013}
\begin{equation}
\label{Eq:69}
	\left(\begin{array}{c}\hat{c}_{0s}\\\hat{c}_{0s}^\dagger\end{array}\right)
=\left(\begin{array}{c}1\\K_1\end{array}\right)e^{-\lambda_1t}
\left(\hat{c}_1+\int\limits_0^te^{\lambda_1t'}\hat{L}_1(t')dt' \right)
+\left(\begin{array}{c}1\\K_2\end{array}\right)e^{-\lambda_2t}
\left(\hat{c}_2+\int\limits_0^te^{\lambda_2t'} \hat{L}_2(t')dt' \right),
\end{equation}
where $\lambda_{1,2}$ and $\left(\begin{array}{c}1\\K_{1,2}\end{array}\right)$ are eigenvalues and eigenvectors of the $2\times2$ matrix:
\begin{eqnarray}
\label{Eq:70}
	\left(\begin{array}{cc}
\Gamma_s&\varsigma\\\varsigma^*&\Gamma_i
\end{array}\right)
\times\left(\begin{array}{c}1\\K_{1,2}\end{array}\right)
=\lambda_{1,2}\left(\begin{array}{c}1\\K_{1,2}\end{array}\right),
\\
\label{Eq:71}
	\left.\begin{array}{cc}
\hat{c}_1=\displaystyle\frac{K_2\hat{c}_s(0)-\hat{c}_i^\dagger(0)}{K_2-K_1},&
\hat{c}_2= - \displaystyle\frac{K_1\hat{c}_s(0)-\hat{c}_i^\dagger(0)}{K_2-K_1}\\
\hat{L}_1=\displaystyle\frac{K_2\hat{L}_s-\hat{L}_i^\dagger}{K_2-K_1},&
\hat{L}_2= - \displaystyle\frac{K_1\hat{L}_s-\hat{L}_i^\dagger}{K_2-K_1}
\end{array}\right\},
\end{eqnarray}
$\hat{c}_s(0)$ and $\hat{c}_i^\dagger(0)$ are initial conditions.

From the solution (\ref{Eq:69})-(\ref{Eq:70}) one can derive a standard-looking condition for the parametric instability(see e.g. \cite{shen}): 
\begin{equation}
\label{Eq:72}
	|\varsigma|^2>\Gamma_s\Gamma_i.
\end{equation}
Consider the inequality (\ref{Eq:72}) in more detail, neglecting for clarity the frequency dispersion of the dielectric filling of a cavity. Taking into account Eq.~(\ref{Eq:56}), one can derive from Eq.~(\ref{Eq:60}) that 
\begin{equation}
\label{Eq:73}
|\varsigma|=|\chi^{(2)}E_p|\sqrt{\omega_s\omega_i}\frac{8\pi l}{L_z\varepsilon}\frac{J}{L_x L_y},
\end{equation}
where the dimensionless factor $J/(L_x L_y)$ depends only on the spatial structure of the modes with frequencies $\omega_{p,s,i}$. From Eqs.~(\ref{Eq:72}), (\ref{Eq:73}) the instability condition takes the form
\begin{equation}
\label{Eq:74}
	\frac{256\pi^2}{\varepsilon^2L_z^2}(\chi^{(2)}l)^2|E_p|^2\left(\frac{\omega_s}{\Delta\omega_s}\right)
\left(\frac{\omega_i}{\Delta\omega_i}\right)\left(\frac{J}{L_x L_y}\right)^2>1,
\end{equation} 
where $\Delta\omega_{s,i}=2\Gamma_{s,i}$ are the linewidths for the signal and idler modes. 

To avoid cumbersome expressions, consider the decay of a pump photon into identical quanta as in Eq.~(\ref{Eq:63}). In this case the instability condition is $|\varsigma|>\Gamma_s$. It is convenient to choose the phase of the pump mode so that the value of  is real and positive. Then Eqs.~(\ref{Eq:69})-(\ref{Eq:71}) yield
\begin{multline}
\label{Eq:75}
	\hat{c}_{0s}=e^{-\Gamma_st}\left[\hat{c}_s(0)\text{cosh}(\varsigma t)-\hat{c}_s^\dagger(0)\text{sinh}(\varsigma t)\right]
+\int\limits_0^te^{(-\varsigma+\Gamma_s)(t'-t)}\frac{\hat{L}_s(t')-\hat{L}_s^\dagger(t')}{2}dt'+\\
+\int\limits_0^te^{(\varsigma+\Gamma_s)(t'-t)}\frac{\hat{L}_s(t')+\hat{L}_s^\dagger(t')}{2}dt'.
\end{multline}
Taking into account the properties of Langevin operators in Eq.~(\ref{Eq:66}) and taking $	\left<\hat{c}_s^\dagger(0)\hat{c}_s^\dagger(0)\right>=\left<\hat{c}_s(0)\hat{c}_s(0)\right>=0$ as an initial state, one can derive from Eq.~(\ref{Eq:75}) the average photon numbers for signal modes $	n_s=\left< \hat{c}_{s}^\dagger\hat{c}_{s}\right>=\left<\hat{c}_{0s}^\dagger\hat{c}_{0s}\right>$:
\begin{multline}
\label{Eq:76}
	n_s=e^{-2\Gamma_st}\left\{n_s(0)\left[ \text{cosh}^2(\varsigma t)+\text{sinh}^2(\varsigma t)\right]+\text{sinh}^2(\varsigma t) \right\}\\
+\left[\Gamma_{\sigma s}n_{T_\sigma}(\omega_s)+\Gamma_{rs}n_{T_r}(\omega_s)\right]
\times\left(\frac{1-e^{2(\varsigma-\Gamma_s)t}}{2(-\varsigma+\Gamma_s)}+\frac{1-e^{2(-\varsigma-\Gamma_s)t}}{2(\varsigma+\Gamma_s)}\right)\\
+\Gamma_s
\left(\frac{1-e^{2(\varsigma-\Gamma_s)t}}{4(-\varsigma+\Gamma_s)}+\frac{1-e^{2(-\varsigma-\Gamma_s)t}}{4(\varsigma+\Gamma_s)}-\frac{1-e^{-2\Gamma_st}}{2\Gamma_s}\right),
\end{multline}
where $\Gamma_s=\Gamma_{\sigma s}+\Gamma_{rs}$. When the parametric amplification starts from the level of vacuum fluctuations, one should put $n_s(0)=0$ in Eq.~(\ref{Eq:76}). 

In the limit of zero pumping intensity, Eq.~(\ref{Eq:76}) gives an expression which describes how the initial perturbation of a photon number approaches equilibrium: 
\begin{equation}
\label{Eq:77}
	n_s=e^{-2\Gamma_st}n_s(0)+\frac{\Gamma_{\sigma s}n_{T_\sigma}(\omega_s)+\Gamma_{rs}n_{T_r}(\omega_s)}{\Gamma_s}\times\left(1-e^{-2\Gamma_st}\right).
\end{equation}

Above the instability threshold, when $\varsigma\gg\Gamma_s$, it is enough to keep only exponentially growing terms in Eq.~(\ref{Eq:76}). We can also assume that an average number of thermal photons in an ambient space $n_{T_r}(\omega_s)$ is negligible. This gives an expression for the parametric signal power $P_s	\approx 2\varsigma\hbar\omega_s n_s$:
\begin{equation}
\label{Eq:78}
	P_s\approx \varsigma\hbar\omega_s e^{2\varsigma t}
\left[ n_s(0)+\frac{\Gamma_{\sigma s}}{\varsigma}n_{T_\sigma}(\omega_s)+\frac{1}{2} \right].
\end{equation}
Obviously this expression is valid only at the initial linear stage. The subsequent evolution is governed by the nonlinear pump depletion and nonlinear modification of phasematching conditions and losses.  An order-of magnitude estimate of the maximum steady-state power  can be obtained from Manley-Rowe relations as shown below for a specific example. 

The fractions of the power emitted outside and absorbed inside a cavity are $P_{r s}\approx\Gamma_{r s}P_s/\varsigma$ and $P_{\sigma s}\approx\Gamma_{\sigma s}P_s/\varsigma$ respectively; most of the power is accumulated in a cavity. From Eq.~(\ref{Eq:78}) one can see that the amplification of intrinsic thermal noise of a QW layer with temperature $T_\sigma$ can be ignored if $\displaystyle\frac{\Gamma_{\sigma s}}{\varsigma}\cdot\frac{2}{\text{exp}(\hbar\omega_s/T_\sigma)-1}\ll1$.

When the parametric growth rate is lower than losses, $\varsigma < \Gamma_s$, the general solution Eq.~(\ref{Eq:76}) describes the regime of spontaneous parametric down-conversion. In the stationary limit, when $(\Gamma_s - \varsigma) t \rightarrow \infty$, the radiated signal power  $P_s \approx 2 \Gamma_{rs} \hbar \omega_s n_s$  becomes 
\begin{equation}
\label{spont}
P_s = \hbar \omega_s \frac{2 \Gamma_{rs} \Gamma_s}{\Gamma_s^2 - \varsigma^2} \left[\Gamma_{\sigma s} n_{T_{\sigma}} (\omega_s) + \Gamma_{rs} n_{T_r} (\omega_s) \right] + \hbar \omega_s \frac{\Gamma_{rs} \varsigma^2}{\Gamma_s^2 - \varsigma^2}.
\end{equation} 
The first term in the right-hand side of Eq.~(\ref{spont}) is due to the thermal emission modified by the parametric decay of the pump photons. The second term originates from the parametric decay of the pump photons under the action of vacuum fluctuations of the intracavity field; this is a purely quantum effect.  Thermal effects can be neglected if 
$$ \frac{\Gamma_s \Gamma_{(\sigma,r)s}}{\varsigma^2} \frac{2}{\exp\left(\hbar\omega_s/(k_B T_{\sigma,r}) \right) -1 } \ll 1. 
$$ 

The results in Eqs.~(\ref{Eq:76})-(\ref{spont}) provide the dependence of the parametric signal from all relevant dissipation and fluctuation effects.  In addition to dissipation and thermal fluctuations due to absorption in the cavity walls and a semiconductor heterostructure, they take into account outcoupling of a signal into the ambient space and coupling to thermal photons from the environment.  Eq.~(\ref{spont}) determines the spontaneous parametric signal emitted from a cavity against the background of noise created by both thermal radiation from a cavity and reemission of thermal photons coupled into a cavity from the outside. The background noise depends from the cavity temperature and the environment temperature. In addition to the spontaneous decay process we take into account the modification of background noise by pumping. 

For a numerical example, consider a nanocavity filled with multiple quantum wells, excited with a mid-infrared pump, as reported in \cite{lee2016}. Using Eq.~(\ref{Eq:74}) for their values of intersubband nonlinearity $|\chi^{(2)}| \sim 3\times  10^{-7}$ m/V, $\omega_{s,i}/\Delta \omega_{s,i} \sim 20$ and $J/L_x L_y \sim 0.3$ we obtain the intracavity pump field at the instability threshold to be $E_p \simeq 8$ kV/cm, which is achievable and is lower than the saturation field for the intersubband nonlinearity. Above the threshold, the signal and idler fields start growing inside the cavity until they get limited by the Manley-Rowe relations \cite{LL}, i.e.~the intracavity signal field reaches $|E_s| \simeq |E_p|/\sqrt{2}$. Therefore, the maximum output signal power that can be obtained per one nanocavity described in \cite{lee2016} is about $8 \times 10^{-7}$ W for the photon leakage rate $\Gamma_{rs} = 10^{12}$ s$^{-1}$. 

Far below the instability threshold, when $\varsigma \ll \Gamma_s$, the spontaneous rate of parametric down-conversion scales as $(\Gamma_{rs}/\Gamma_s^2) \varsigma^2$. For the parameters from the above numerical example, when $\varsigma = \Gamma_s/2$ the emission rate of signal photons is around $3\times 10^{11}$ s$^{-1}$ and the power is 6 nW. 

A very high second-order nonlinear surface conductivity for graphene was reported in \cite{wang2016}, corresponding to the effective bulk susceptibility $|\chi^{(2)}| \sim 10^{-3}$ m/V per monolayer in the THz range. This large susceptibility if partially offset by a small factor $l/L_z$ in Eq.~(\ref{Eq:73}) where $l$ is a thickness of the graphene layer. However, for hBN-encapsulated graphene utilized to fabricate low-disorder graphene samples \cite{low2016} the total cavity thickness $L_z$ can be as small as several nm, so the factor  $l/L_z$ can be as large as 0.1. Even factoring in enhanced cavity losses, this can yield a lower parametric instability threshold as compared to semiconductor quantum well samples.  

\section{Comparing parametric instability in a subwavelength cavity and in a Fabry-Perot cavity  }

Compare the parametric instability in a subwavelength cavity with similar instability of modes in a Fabry-Perot (FP) cavity with all three dimensions larger than wavelength, which we will call a quasi-optical cavity. Consider a planar quasi-2D cavity of the surface area $L_x L_y$, in which the waves are propagating along the nonlinear layer of thickness $l$ much smaller than the cavity thickness $L_{FP}$ transverse to the nonlinear layer, so the cavity volume is $L_{FP} L_x L_y$. The dielectric constant of a cavity filling is $\varepsilon$.  In this case the parametric down-conversion is still described by Eqs.~(\ref{Eq:66}), (\ref{Eq:60}), and (\ref{Eq:61}), in which the relaxation constants $\Gamma_{s,i}$ and the overlap integral are determined by the FP cavity Q-factor and the corresponding spatial structure of the modes. For the normalization constants of the quantum fields entering Eq.~(\ref{Eq:60}) we use standard expressions for a two-mirror FP cavity: 
\begin{gather}
\label{Eq:79}
	|E_{s,i}|^2 = \frac{4 \pi \hbar \omega_{s,i}}{L_x L_y L_{FP} \varepsilon}.  
\end{gather} 
The resulting parametric instability threshold is 
\begin{equation}
\label{Eq:80}
	\frac{64\pi^2}{\varepsilon^2L_{FP}^2}\left(\chi^{(2)}l\right)^2|E_p|^2 \left(\frac{\omega_s}{\Delta \omega_s} \right)  \left(\frac{\omega_i}{\Delta \omega_i} \right)  \left(\frac{J}{L_x L_y} \right)^2  >1,
\end{equation}
where $\Delta \omega_{s,i} \approx 2 \Gamma_{s,i}$. 

As we already pointed out, in a cavity with all three dimensions larger than the wavelength the phase matching condition for a three-wave mixing may be more difficult to satisfy. Even if we assume that phase matching is somehow arranged and the geometric factor $ J/L_x L_y$ is of the same order as in a subwavelength cavity, the latter is expected to have a lower parametric threshold. Indeed the ratio of the threshold pump intensity $|E_p|^2$ in a subwavelength cavity to that in a quasi-optical cavity scales as $\sim \displaystyle \left(\frac{L_z}{L_{FP}}\right)^2\left(\frac{\Delta\omega_{eff}}{\Delta\omega_{FP}}\right)^2$, where $\Delta\omega_{eff}$ and $\Delta\omega_{FP}\approx\Delta\omega_{s,i}$ are typical linewidths of the subwavelength cavity and FP cavity modes, respectively. This ratio can be much smaller than 1, which indicates that a much lower pumping is needed to reach the parametric  instability threshold in a subwavelength cavity, even if the FP cavity has a higher Q-factor as compared to the subwavelength cavity, $ \Delta\omega_{FP} < \Delta\omega_{eff}$.  

A plane-parallel quasi-2D subwavelength cavity geometry considered in this paper is the most natural choice for integration with 2D nonlinear materials. However, other geometries are also possible, for example plasmonic or grating structures supporting surface modes. To get an order of magnitude estimate of the parametric threshold, one can use our results in Eqs.~(\ref{Eq:74}), (\ref{Eq:80}) after replacing $L_z$ or $L_{FP}$ with a mode size transversely to the nonlinear layer. 

A promising example of such a plasmonic nanocavity was reported in \cite{akselrod2015}. It consists of a monolayer MoS$_2$, which is a 2D semiconductor, sandwiched between a gold substrate and a patch silver nanoantenna. Such a cavity has high radiative and absorption losses but a very small transverse mode size of less than 10 nm and an ultrasmall effective mode volume of $ \sim 10^{-3} (\lambda/\sqrt{\epsilon})^3$. The authors of \cite{akselrod2015} used their cavities to obtain a 2000-fold enhancement in the photoluminescence intensity from MoS$_2$ monolayer. However, a cavity of similar design can also be used for parametric down-conversion from visible to the near-IR  range. A high second-order nonlinearity, about an order of magnitude higher than in BBO or lithium niobate, has been reported for MoS$_2$ monolayer \cite{heinz2013}. An even higher nonlinearity has been observed in the vicinity of exciton resonances \cite{pedersen2015}. Assuming conservatively that the effective second-order susceptibility for MoS$_2$ is $|\chi^{(2)}| \sim 10^{-10}$ m/V, monolayer thickness 0.6 nm, transverse mode size 5 nm, $\omega_{s,i}/\Delta \omega_{s,i} \sim 20$ and $J/L_x L_y \sim 0.3$ we obtain from Eq.~(\ref{Eq:74}) the intracavity pump field at the parametric amplification threshold to be $E_p \simeq 30$ MV/cm, which is much higher than the estimate above for a nonlinear cavity based on mid-infrared resonant intersubband nonlinearity of quantum wells, but is below the saturation threshold for MoS$_2$ and easily achievable with pulsed lasers. 

Ultracompact subwavelength electrodynamic structures utilizing 2D materials are promising for applications in integrated photonic circuits, whenever one needs a compact planar architecture. At the same time, due to strong dissipation they are unlikely to outperform conventional nonlinear devices made of bulk transparent nonlinear materials when it comes to the nonlinear conversion efficiency and power. For example, in \cite{furst2010} the authors realized low-threshold mode-matched parametric generation in whispering gallery mode resonators made entirely of bulk lithium niobate.  In this case  the bulk nonlinear material occupies all modal volume. The lower nonlinearity and the loss of Purcell enhancement are  compensated by lower dissipation and increased interaction volume.

In conclusion, we applied a consistent Heisenberg-Langevin formalism to the process of nonlinear parametric down-conversion  of cavity modes in planar subwavelength cavities containing 2D nonlinear materials. We derived general analytic formulas for the spontaneous parametric signal and  threshold of stimulated parametric down-conversion of a pump cavity mode into the signal and idler modes. We found that a significant reduction in the parametric instability threshold  can be achieved for realistic materials and cavity parameters due to Purcell enhancement. 

This material is based upon work supported by the Air Force Office of Scientific Research under award numbers FA9550-17-1-0341 and FA9550-15-1-0153.  M.T. acknowledges the support from RFBR grant No. 17-02-00387 and from the program of the Presidium of the Russian Academy of Sciences ``Nonlinear dynamics in mathematical and physical sciences'' (project No. 0035-2018-0006).

\appendix

\section{Commutation relations for Langevin sources  \label{appendix:Commutation relations}}

Consider first a single quantum oscillator described by the Hamiltonian $\hat{H}= \hbar \omega(\hat{c}^\dagger \hat{c} + 1/2) $. After substituting $\hat{c}= \hat{c}_0 e^{-i\omega t}$ and $\hat{c}^\dagger= \hat{c}^\dagger_0 e^{-i\omega t}$  the Heisenberg equations of motion take the form $\dot{\hat{c}}_0 = 0$, $\dot{\hat{c}}^\dagger_0 = 0$. The simplest model of interaction with a dissipative reservoir modifies these equations as follows: $\dot{\hat{c}}_0  + \Gamma \hat{c}_0 =0$, $\dot{\hat{c}}^\dagger_0  + \Gamma \hat{c}^\dagger_0 =0$. However, this modification leads to violation of boson commutation relation $[\hat{c}_0 ,\hat{c}^\dagger_0]=1$. To resolve this issue and preserve the commutator one has to add the Langevin sources to the right-hand side of Heisenberg equations \cite{SZ,gardiner,tokman2018}:

\begin{equation}
\label{Eq:C1}
\dot{\hat{c}}_0  + \Gamma \hat{c}_0 = \hat{L},
\hspace{.4cm}
\dot{\hat{c}}^\dagger_0  + \Gamma \hat{c}^\dagger_0 = \hat{L}^\dagger.
\end{equation}
Langevin noise operators in  Eq.~(\ref{Eq:C1}) describe fluctuations in a dissipative system. Note that $\langle\hat{L}\rangle=0$ ; the notation $\langle\cdots\rangle$ means averaging over the statistics of the dissipative reservoir and over the initial quantum state $|\Psi \rangle$ within the Heisenberg picture.

The  commutation relations for a noise operator can be obtained directly from the given form of the relaxation operator if we require that standard commutation relations $[\hat{c}_0 ,\hat{c}^\dagger_0]=1, [\hat{c}_0 ,\hat{c}_0]=0$,   be satisfied at any moment of time \cite{gardiner,tokman2018}. Indeed, let's substitute the solution of the operator-valued equations~(\ref{Eq:C1})

\begin{equation}
\label{Eq:C2}
\hat{c}_0 = \hat{c}_0(0) e^{-\Gamma t} + \int\limits_0^t e^{ \Gamma (t'-t) } \hat{L}(t') dt',
\hspace{.4cm}
\hat{c}^\dagger_0 = \hat{c}^\dagger_0(0) e^{-\Gamma t} + \int\limits_0^t  e^{ \Gamma (t'-t) } \hat{L}^\dagger(t') dt'
\end{equation}
into the commutators. It is easy to see that the standard commutation relations will be satisfied if, first of all, the field operators at an initial moment of time, $\hat{c}_0 (0)$ and $\hat{c}^\dagger_0(0)$, commute with Langevin operators $\hat{L}(t)$ and $\hat{L}^\dagger(t)$ in any combination. Second, the following condition has to be satisfied:  

\begin{equation}
\label{Eq:C3}
[\hat{L} ,\hat{c}^\dagger_0]=[\hat{c}_0,\hat{L}^\dagger] = \Gamma.
\end{equation}
Substituting  Eq.~(\ref{Eq:C2}) into  Eq.~(\ref{Eq:C3}) and using the identity $\int\limits_0^t  X(t') \delta(t-t') dt' = X(t)/2$  we arrive at

\begin{equation}
\label{Eq:C4}
[\hat{L}(t') , \hat{L}^\dagger(t)] = 2\Gamma \delta(t-t') .
\end{equation}

Now consider an ensemble of coupled oscillators  Eq.~(\ref{Eq:66}). One can find directly from the solution  Eq.~(\ref{Eq:69}) that the following conditions have to be satisfied in order to preserve standard commutation relations $[\hat{c}_{0s} ,\hat{c}^\dagger_{0s}]=[\hat{c}_{0i} ,\hat{c}^\dagger_{0i}] = 1 , [\hat{c}_{0s} ,\hat{c}_{0i}]=0
$ etc.:

\begin{equation}
\label{Eq:C5}
\left.
\begin{matrix}

[\hat{L}_s ,\hat{c}^\dagger_{0s}]=[\hat{c}_{0s} , \hat{L}^\dagger_s] = \Gamma_s\\
[\hat{L}_i ,\hat{c}^\dagger_{0i}]=[\hat{c}_{0i} , \hat{L}^\dagger_i] = \Gamma_i\\\
[\hat{L}_s ,\hat{c}^\dagger_{0i}]=[\hat{c}_{0s} , \hat{L}^\dagger_i] = [\hat{L}_s , \hat{c}_{0i}]= [ \hat{c}_{0s}, \hat{L}_i] = 0
\end{matrix}
\right\}.
\end{equation}
It is easy to find out that  Eqs.~(\ref{Eq:C5}) will be satisfied if the field operators at $t=0$ commute with Langevin noise operators in any combination, and the noise operators $\hat{L}_s$ and $\hat{L}_i$ commute with each other. In addition, substituting Eq.~(\ref{Eq:69}) -~(\ref{Eq:71}) into  Eq.~(\ref{Eq:C5}) one can show that in order to satisfy  Eq.~(\ref{Eq:C5}) the following relations must hold: 

\begin{equation}
\label{Eq:C6}
\displaystyle  \frac{ \int\limits_0^t  \left( K_2 e^{\lambda_1 (t'-t)} - K_1 e^{\lambda_2 (t'-t)} \right)\left[ \hat{L}_s(t'),\hat{L}^\dagger _s (t)\right]dt' }{K_2 - K_1} = \Gamma_s,
\end{equation}
\begin{equation}
\label{Eq:C7}
\displaystyle  \frac{ \int\limits_0^t  \left( K_2 e^{\lambda_2 (t'-t)} - K_1 e^{\lambda_1 (t'-t)}\right) \left[ \hat{L}_i(t'),\hat{L}^\dagger _i (t)\right]dt' }{K_2 - K_1} = \Gamma_i.
\end{equation}
From  Eqs.~(\ref{Eq:C6}) and~(\ref{Eq:C7}) one can obtain the requirement Eq.~(\ref{Eq:67}) which preserves correct commutators of the field operators. 
Therefore, the commutation properties of correct noise operators for coupled oscillators have to be exactly the same as for uncoupled isolated oscillators. 

Here we presented a general proof which does not rely on any specific microscopic model of a dissipative subsystem coupled to the field oscillators.  The proof for a particular case of two identical coupled oscillators interacting with a standard dissipative reservoir of equilibrium harmonic oscillators \cite{SZ} has been recently obtained in \cite{martynov2017}.


\begin{thebibliography}{30}

\bibitem{purcell}E. M. Purcell, H. C. Torrey, and R. V. Pound, Phys. Rev. 69, 37 (1946).




\bibitem{lin1994} H.-B. Lin and A. J. Campillo, Phys. Rev. Lett. 73, 2440 (1994). 
\bibitem{bermel2007} P. Bermel, A. Rodriguez, J. D. Joannopoulos, and M. Soljacic, Phys. Rev. Lett. 99, 053601 (2007). 
\bibitem{genevet2010} P. Genevet, J.-P. Tetienne, E. Gatzogiannis, R. Blanchard,
M. A. Kats, M. O. Scully, and F. Capasso, Nano Lett. 10, 4880 (2010). 

\bibitem{kauranen2012} M. Kauranen  and A. V. Zayats, Nature Phot. 6, 737 (2012). 
\bibitem{kivshar2014} M. Lapine, I. V. Shadrivov, and Y. S. Kivshar, Rev. Mod. Phys. 86, 1093 (2014). 

\bibitem{lee2014} J. Lee et al., Nature 511, 65 (2014). 
\bibitem{lee2016} J. Lee et al.,  Adv. Opt. Mater. 4, 664 (2016). 

\bibitem{smirnova2016} D. Smirnova and Y. S. Kivshar, Optica 3, 1241 (2016). 
\bibitem{atwater2017} A. R. Davoyan and H. A. Atwater, Optica 5, 608 (2018). 

\bibitem{hendry2010} E. Hendry, P.J. Hale, J. Moger, A.K. Savchenko, and S.A. Mikhailov, PRL 105, 097401 (2010).
\bibitem{yao2012} X. Yao and A. Belyanin, Phys. Rev. Lett. 108, 255503 (2012).
\bibitem{yao2014}	X. Yao, M.D. Tokman, and A. Belyanin, Phys. Rev. Lett. 112, 055501 (2014).
\bibitem{wang2016} Y. Wang, M. Tokman, and A. Belyanin,  Phys. Rev. B 94, 195442 (2016).
\bibitem{malard2013} L.M. Malard, T.V. Alencar, A.P.M. Barboza, K.F. Mak, and A.M. de Paula,  Phys. Rev. B 87, 201401 (2013).
\bibitem{liu2016} H. Liu, Y. Li, Y. S. You, S. Ghimire, T. F. Heinz, and D. A. Reis, Nat. Phys. 13, 262 (2016).
\bibitem{koppens2016} P. Alonso-Gonz‡lez, et al., Nat. nanotechnol. 12, 31 (2016). 
\bibitem{ye2015} Yu Ye, Z. Wong, X. Lu, X. Ni, H. Zhu, X.Chen, Y. Wang, 
and X. Zhang, Nat. Phot. 9, 733 (2015). 

 
\bibitem{SZ} M. O. Scully and M. S. Zubairy, Quantum Optics (Cambridge Univ. Press, Cambridge, 1997).

\bibitem{gardiner} C. Gardiner and P. Zoller, Quantum Noise (Springer-Verlag, Berlin, Heidelberg, 2004). 

\bibitem{et2017} M. Erukhimova and M.Tokman. Phys. Rev. A 95, 013807 (2017).
\bibitem{david} L. Davidovich, Rev. Mod. Phys. 68, 127 (1996).
\bibitem{belyanin2013} M.Tokman, X. Yao, and A. Belyanin, Phys. Rev. Lett. 110, 077404 (2013).

\bibitem{et2015} M. Erukhimova, and M. Tokman. Optics Lett. 40,  2739 (2015).
\bibitem{te2013} M. Tokman, M. Erukhimova,  Journal of Lumin. 137, 148 (2013). 
\bibitem{tokman2018} M. Tokman, Z. Long, S. AlMutairi, Y. Wang, M. Belkin, and A. Belyanin, Phys. Rev. A 97, 043801 (2018). 



\bibitem{fain} V. M. Fain and Ya. I. Khanin, Quantum Electronics, Vol. 1 (The MIT Press, Cambridge, MA, 1969).
\bibitem{vdovin} M. D. Tokman, M. A. Erukhimova, and V. V. Vdovin, Annals of Physics 360, 571 (2015).
\bibitem{tokman2016} M. Tokman, Y. Wang, I. Oladyshkin, A. R. Kutayiah, and A. Belyanin, Phys. Rev. B 93, 235422 (2016). 


\bibitem{keldysh} Yu. A. Illinskii and L. V. Keldysh, Electromagnetic Response of Material Media (Springer, New York, 1994).
\bibitem{te2011} M. D. Tokman and M. A Erukhimova, Phys. Rev. E 84, 056610 (2011).

\bibitem{shen} Y. R. Shen, The Principles of Nonlinear Optics (Wiley, Hoboken, NJ, 2003). 

\bibitem{LL} L. D. Landau and E. M. Lifshitz, Electrodynamics of Continuous Media (Pergamon Oxford, 1984). 

\bibitem{low2016} T. Low, A. Chaves, J. D. Caldwell, A. Kumar, N. X. Fang,
P. Avouris, T. F. Heinz, F. Guinea, L. Martin-Moreno, and F. Koppens. Nat. Mater. 16, 182 (2016). 

\bibitem{akselrod2015} G. M. Akselrod, T. Ming, C. Argypoulos, T. B. Hoang, Y. Lin, X. Ling, D. R. Smith, J. Kong, and M. H. Mikkelsen, Nano Lett. 15, 3578 (2015). 

\bibitem{heinz2013} Y. Li, Yi Rao, K. F. Mak, Y. You, S. Wang, C. R. Dean, and T. F. Heinz, Nano Lett. 13, 3329 (2013). 

\bibitem{pedersen2015} M. L. Trolle, Y.-C. Tsao, K. Pedersen, and T. G. Pedersen, Phys. Rev. B 92, 161409(R) (2015). 

\bibitem{furst2010} J. U. Furst, D.V. Strekalov, D. Elser, A. Aiello, U. L. Andersen, Ch. Marquardt, and G. Leuchs,        Phys. Rev. Lett. 105, 263904 (2010).

\bibitem{martynov2017} V. O. Martynov, V A Mironov, and L A Smirnov, J. Phys. B: At. Mol. Opt. Phys. 50,  085501 (2017). 


\end{thebibliography}
\end{document}